\documentclass[12pt]{article}
\usepackage{epsfig}
\usepackage{pstricks,rotating, graphicx}
\usepackage{a4}
\usepackage{latexsym}
\usepackage{cite}

\usepackage{color}
\usepackage{colordvi}
\usepackage[hang]{subfigure}

\graphicspath{{Pics/}}
\DeclareGraphicsExtensions{.eps,.ps}

\usepackage[width=16.5cm, left=2.2cm,top=2.5cm,height=24.cm]{geometry}

\usepackage{pslatex}
\usepackage{txfonts}
\usepackage[latin1]{inputenc}
\usepackage[T1]{fontenc}

\newcommand{\GeV}{\ensuremath{\,\mathrm{GeV}}}

\def\muf{{\mu^{}_f}}
\def\mufs{{\mu^{\,2}_f}}
\def\mur{{\mu^{}_r}}
\def\murs{{\mu^{\,2}_r}}
\def\alphas{{\alpha_s}}

\def\MSbar{{$\overline{\mbox{MS}}\,$}}
\def\mbar{\overline{m}}
\def\mmu{m(\mur)}
\def\mm{m(m)}

\begin{document}

\begin{titlepage}
\noindent
DESY 10-212 
\vspace{1.3cm}

\begin{center}
  {\bf
    \Large
    Heavy-quark deep-inelastic scattering with a running mass \\
  }
  \vspace{1.5cm}
  {\large
    S.~Alekhin$^{\, a,b}$\footnote{{\bf e-mail}: sergey.alekhin@ihep.ru}
    and S.~Moch$^{\, a,}$\footnote{{\bf e-mail}: sven-olaf.moch@desy.de} \\
  }
  \vspace{1.2cm}
  {\it
    $^a$Deutsches Elektronensynchrotron DESY \\
    Platanenallee 6, D--15735 Zeuthen, Germany \\
    \vspace{0.2cm}
    $^b$Institute for High Energy Physics \\
    142281 Protvino, Moscow region, Russia\\
  }
  \vspace{1.4cm}
  \large {\bf Abstract}
  \vspace{-0.2cm}
\end{center}
We study the production of heavy quarks in deep-inelastic scattering within perturbative QCD. 
As a novelty, we employ for the first time the running mass definition in the \MSbar scheme 
for deep-inelastic charm and bottom production.
We observe an improved stability of the perturbative expansion and 
a reduced theoretical uncertainty due to variations of 
the renormalization and factorization scales.
As our best estimate we extract from a global fit to fixed-target and HERA collider data 
for the charm-quark an \MSbar mass of $m_c(m_c)$ = 1.01 $\pm$ 0.09 (exp)  $\pm$ 0.03 (th) GeV.

\end{titlepage}

\section{Introduction}
\label{sec:intro}

The production of heavy quarks in deep-inelastic scattering (DIS) 
is an important reaction and has been measured with high accuracy 
in several fixed-target experiments and at the HERA collider.
Within perturbative QCD, the production of charm and bottom quarks proceeds
in neutral (NC) or charged current (CC) reactions 
via lepton-parton scattering and the exchange of a virtual boson 
$\gamma^*/Z$ or $W^\pm$ with space-like momentum.
A detailed understanding of the production mechanism sheds light on the 
underlying parton dynamics in QCD. 
In the LHC era this, perhaps, is the most important aspect, 
because DIS heavy-quark production provides core constraints in global fits   
on the parton distribution functions (PDFs) even at the terascale. 

Thus, it is of paramount importance to provide precision predictions 
which, of course, have to rely on higher order radiative corrections. 
Our theory predictions for heavy-quark production include next-to-leading order (NLO) 
QCD corrections~\cite{Laenen:1992zk,Gottschalk:1980rv,Gluck:1996ve} and, 
in the case of NC DIS, even partial information at next-to-next-to-leading order (NNLO), 
which comprises in particular all logarithmically enhanced terms near threshold~\cite{Presti:2010pd}, 
and all explicit dependence on the renormalization and factorization scales.

However, precision predictions, must also address the uncertainty due to the non-perturbative parameters, 
such as the aforementioned PDFs, the value of the strong coupling constant $\alphas$ and the mass $m$ of the heavy quarks 
charm, bottom and top. 
It is precisely with respect to the latter aspect, that we wish to improve the current state-of-the-art. 
Namely, we employ the short-distance (so-called \MSbar) mass in our treatment of heavy-quark DIS.
In this manner we provide a crucial link, which has long been missing, 
for the comparison of the heavy-quark masses entering in DIS 
and the determination of PDFs in global fits on the one hand, 
and, on the other, those obtained from other determinations, 
e.g. in $e^+e^-$-collisions or by means of lattice computations.

Traditionally, perturbative corrections to hard scattering processes at hadron colliders 
have used the so-called pole mass of the heavy quark as a definition of the mass parameter.
The pole mass is popular, because it is well defined at each finite order of perturbation theory 
and it is introduced in a gauge invariant way.
However, as is well-known since long, the concept of the pole mass in QCD has intrinsic theoretical limitations.
Because of confinement no free colored quarks exist, 
i.e. they do not appear as an asymptotic state of the $S$-matrix.
It has been shown that the use of the pole mass leads to a poorly behaved perturbative series~\cite{Bigi:1994em}, 
because observables in hard scattering processes become sensitive to momentum regions of the order of the QCD scale $\Lambda_{\rm QCD}$.

Alternative mass definitions offer a solution to this problem. 
The most prominent example is the \MSbar mass $m(\mur)$, 
which is to be evaluated at the (renormalization) scale $\mur$, where $\mur \gg \Lambda_{\rm QCD}$, 
and which is free of ambiguities of order $\Lambda_{\rm QCD}$.
For inclusive cross sections at short distances the appropriate scale choice for the running mass $m(\mur)$ is $\mur=m$, 
where the renormalization group evolution for the scale dependence of the mass converges even for scales as low as the 
charm-quark mass.
As a benefit of theory predictions using the \MSbar mass one observes 
an improved stability of the perturbative series with respect to scale variations 
as compared to the result in the pole mass scheme.

In this study, we start off with QCD predictions for heavy-quark DIS 
at NLO~\cite{Gottschalk:1980rv,Gluck:1996ve,Laenen:1992zk} 
and approximate NNLO~\cite{Presti:2010pd}, which have been computed with a pole mass.
Subsequently, we improve the perturbative description by converting from the pole mass scheme 
to the \MSbar scheme (see Refs.~\cite{Gray:1990yh,Chetyrkin:1999qi,Melnikov:2000qh} and references therein).
The necessary scheme transformation follows closely similar recent work for top-quark production at hadron colliders, 
cf.~\cite{Langenfeld:2009wd,Aliev:2010zk} for implementation details.
From a global fit of the parton distribution functions to fixed-target and
HERA collider data we extract for the first time an \MSbar mass for the charm-quark $m_c(m_c)$.
Our best estimate for $m_c(m_c)$ is consistent with the world average~\cite{Nakamura:2010pdg} 
within the quoted range of errors.
Moreover, PDFs determinations can benefit from reduced uncertainties 
due to precise heavy-quark masses for charm and bottom, $m_c$ and $m_b$, and  
we comment on the implications for $W^\pm$ and $Z$ gauge boson production at the LHC.

\section{Heavy-quark DIS in perturbative QCD}
\label{sec:pQCD}

For NC DIS pair-production at leading order (LO) proceeds through 
photon-gluon fusion as
\begin{equation}
\label{eq:ncborn}
g(p) + \gamma^*(q) \to  q_2 + {\bar q}_2
\, ,
\end{equation}
which is a $2 \to 2$ process starting off at order $\alphas$ in QCD and, 
of course, involving the overall power $\alpha$ for the QED coupling and the 
quark fractional charges.

For CC DIS, on the contrary, heavy-quark production at parton level proceeds in Born approximation 
in a $2 \to 1$ reaction as
\begin{equation}
\label{eq:ccborn}
q_1(p) + W^*(q) \to  q_2
\, ,
\end{equation}
where the initial quark $q_1$ is light, the final state quark $q_2$ is heavy
and the coupling to the $W$-boson involves the usual parameters 
of the Cabibbo-Kobayashi-Maskawa (CKM) matrix.

The well-known kinematical variables are Bjorken $x$ and $Q^2$ 
defined by the momenta $p$ and $q$ of the incoming parton and the off-shell boson,
\begin{equation}
  \label{eq:xQ2def}
  Q^2 \, = \, -q^2 \, > \, 0 \, 
  \qquad\qquad\qquad
  x \, = \, {Q^2 \over 2 p \cdot q }
\, ,
\end{equation}
and the cross sections is conveniently parametrized in terms of 
the heavy-quark DIS structure functions $F_k$, $k=1,2,3$, which depend on 
$x$, $Q^2$ and the heavy-quark mass $m$.
In the standard factorization approach to perturbative QCD 
the structure functions $F_k$ can be written as a convolution of PDFs and
coefficient functions, 
\begin{equation}
  \label{eq:totalF2c}
  F_k(x,Q^2,m^2) =
  \sum\limits_{i = q,{\bar{q}},g} \,\,
  \int\limits_{x}^{z^{\rm max}}\,
  {dz \over z} \,\, f_{i}\left({x \over z}, \mufs \right)\,\,
  {\cal C}_{k, i}\left(z,\xi,\murs,\mufs \right)
  \, ,
\end{equation}
where the renormalization scale is denoted as $\mur$ 
and the PDFs for the parton of flavor $i$  at the factorization scale $\muf$ as $f_{i}(x,\mufs)$.
Depending on the kinematics in Eqs.~(\ref{eq:ncborn}), (\ref{eq:ccborn}),  
the integration range over the parton momentum fraction $z$ extends to 
$z_{\rm (CC)}^{\rm max} = 1/(1 +   m^2/Q^2)$ or 
$z_{\rm (NC)}^{\rm max} = 1/(1 + 4 m^2/Q^2)$.
The kinematical variable $\xi$ in Eq.~(\ref{eq:totalF2c}) is given as 
\begin{equation}
  \label{eq:xi-def}
  \xi = {Q^2 \over m^2}\, .
\end{equation}

The coefficient functions ${\cal C}_{k, i}$ of the hard parton scattering process in Eq.~(\ref{eq:totalF2c}) 
can be computed in a perturbative expansion in the strong coupling constant $\alphas = \alphas(\mur)$. 
Currently, we have for both cases, NC and CC, the complete NLO corrections available 
with full dependence on the heavy-quark mass $m$, see Refs.~\cite{Laenen:1992zk} and ~\cite{Gottschalk:1980rv,Gluck:1996ve}, 
which we use in our description of the heavy-quark structure functions $F_k$.
Specifically, in the NC case, we use the code of Ref.~\cite{Riemersma:1994hv} (see~\cite{Harris:1995tu} for minor corrections).

Beyond NLO, partial results are available, although the complete NNLO corrections are not known to date.
In the asymptotic limit $m^2/Q^2 \to 0$ fully analytic results have obtained at NLO, see \cite{Buza:1995ie,Buza:1997mg,Bierenbaum:2007qe}
and at NNLO for the lowest even-integer Mellin moments \cite{Bierenbaum:2009mv}.
For parton energies close to the production threshold, $s \simeq 4 m^2$, 
soft gluon improvements at NNLO are long known to be important~\cite{Laenen:1998kp,Alekhin:2008hc}. 
For NC heavy-quark production the convolution of the coefficient functions (especially ${\cal C}_{2, g}$) and the gluon density is dominated by rather 
low partonic of-mass energies $s$ and the corresponding soft logarithms in
$\beta = (1 - 4 m^2/s)^{1/2}$ at NNLO have recently been completely determined~\cite{Presti:2010pd}.
In our description of the NC structure functions $F_k$ 
(which supersedes our earlier studies~\cite{Alekhin:2008hc}) 
we include these latest improvements~\cite{Presti:2010pd} together with complete dependence on 
the renormalization and factorization scales, see e.g.~\cite{Laenen:1998kp,Langenfeld:2009wd}.
This approximation to NC DIS we call NNLO$_{\rm approx}$.
Soft logarithms have also been studied for the CC case, see~\cite{Corcella:2003ib}.
However, in the kinematical range of the currently available CC DIS data, 
they are numerically less important and we do not include them here.
Hence, for CC DIS, we confine ourselves to NLO accuracy only.

The mass parameter in the structure functions $F_k$ in Eq.~(\ref{eq:totalF2c}) is the pole mass of the heavy quark, 
which requires $m$ to coincide with the pole of the heavy-quark propagator at each finite order in perturbation theory.
In this way, that value of the mass itself is strongly depended on the perturbative order. 
Moreover, it has intrinsic uncertainties of order $\Lambda_{\rm QCD}/m$. 
The perturbative description of heavy-quark DIS can be improved, however, by
performing a scheme change from the pole mass to the \MSbar scheme. 

The starting point of this conversion is the well-known relation between the pole mass $m$ 
and the running mass $\mmu$ in the \MSbar scheme
\begin{equation}
  \label{eq:mpole-mbar}
  m = \mmu \* \left(1 + \alphas(\mur) d^{(1)}(\mur) + \alphas(\mur)^2 d^{(2)}(\mur) + \dots \right)
  \, ,
\end{equation}
where the coefficients $d^{(l)}$ of the  perturbative expansion in $\alphas$
are actually known to three-loop order~\cite{Gray:1990yh,Chetyrkin:1999qi,Melnikov:2000qh}.

\bigskip

Let us start with the NC case. 
We will derive explicit formulae through NNLO for the dependence of the structure functions on the \MSbar mass $\mm$.
In doing so, we follow similar recent work for the pair-production of top-quarks at hadron colliders~\cite{Langenfeld:2009wd,Aliev:2010zk}.
For the pole mass $m$ we have (suppressing all other arguments),
\begin{equation}
  \label{eq:F2c-mpole}
  F_k(m) =
  \alphas\, F_k^{(0)}(m) +  \alphas^2\, F_k^{(1)}(m) +
  \alphas^3\, F_k^{(2)}(m) 
  \, ,
\end{equation}
which we can convert with Eq.~(\ref{eq:mpole-mbar}) to the \MSbar mass
$\mm$ (for simplicity abbreviated as $\mbar$) according to 
\begin{eqnarray}
  \label{eq:F2c-mbar}
  F_k(\mbar) &=&
  \alphas\, F_k^{(0)}(\mbar) 
\\
& &
\nonumber
  + 
  \alphas^2\, \left( 
    F_k^{(1)}(\mbar) 
    + \mbar\, d^{(1)} \partial_m F_k^{(0)}(m) \biggr|_{m=\mbar}
  \right) 
\\
& &
\nonumber
  +
  \alphas^3\, \left( 
     F_k^{(2)}(\mbar) 
     + \mbar\, d^{(2)} \partial_m F_k^{(0)}(m) \biggr|_{m=\mbar}
     + \mbar\, d^{(1)} \partial_m F_k^{(1)}(m) \biggr|_{m=\mbar}
\right.
\\
& &
\nonumber
\left.
\hspace*{60mm}
     + {1 \over 2}\, \left(\mbar\, d^{(1)}\right)^2 \partial_m^2 F_k^{(0)}(m) \biggr|_{m=\mbar}
   \right) 
  \, ,
\end{eqnarray}
where the coefficients $d^{(l)}$ have to be evaluated for $\mur = \mbar$ (corresponding to the scale of $\alphas$).

In the NC case, the coefficient functions in Eq.~(\ref{eq:totalF2c}) 
have a perturbative expansion to NNLO in the strong coupling $\alphas = \alphas(\mur)$,
\begin{equation}
  \label{eq:partonCnc-exp}
  {\cal C}_{k, i}(\eta(z),\xi,\murs,\mufs) =
  \alphas\, {\cal C}_{k, i}^{(0)} +  \alphas^2\, {\cal C}_{k, i}^{(1)} +
  \alphas^3\, {\cal C}_{k, i}^{(2)}
  \, ,
\end{equation}
where $\eta$ denotes the distance to partonic threshold. 
The partonic center-of-mass energy reads $s=Q^2 (1/z-1)$, so that
\begin{equation}
  \label{eq:eta-def}
  \eta(z) 
  \,=\, {s \over 4 m^2} - 1
  \,=\, {Q^2 \over 4 m^2}\left({1 \over z} -1 \right) - 1
  \, .
\end{equation}

Mass dependence resides in the coefficient functions implicitly 
in $\eta$ and $\xi$ as well as in the factorization scale dependent part 
(commonly appearing through the ratio $\mufs/m^2$).
Thus
\begin{equation}
  \label{eq:cdiff}
  \partial_m {\cal C}_{k, i}(\eta(z),\xi,\murs,\mufs) \,=\,
  \left( \partial_m \eta \right)\, \partial_\eta {\cal C}_{k, i}
  +
  \left( \partial_m \xi \right)\, \partial_\xi {\cal C}_{k, i}
  - 
  {\muf \over m} \partial_\muf {\cal C}_{k, i}
\, .
\end{equation}
In this way the explicit expression for the first-order derivative $\partial_m$ in
Eq.~(\ref{eq:F2c-mbar}) becomes 
\begin{eqnarray}
\label{eq:partial-F2c}
  \partial_m F_k^{(l)}(m) &=&
  \sum\limits_{i = q,{\bar{q}},g} \,\,
  \int\limits_{x}^{z^{\rm max}}\,
  {dz \over z} \,\, f_{i}\left({x \over z}\right)\,
\left\{
2 {z (1-z) \over m}\, \partial_z {\cal C}_{k, i}^{(l)}
- {2 \over m} \xi\, \partial_\xi {\cal C}_{k, i}^{(l)}
- {\muf \over m} \partial_\muf {\cal C}_{k, i}^{(l)}
\right\}
\\
& &
  + 
\left( \partial_m z^{\rm max} \right)\, 
  \sum\limits_{i = q,{\bar{q}},g} \,\,
  {1\over z}\, f_{i}\left({x \over z}\right)\,
    {\cal C}_{k, i}^{(l)}
    \biggr|_{z=z^{\rm max}}
    \, ,
\nonumber
\end{eqnarray}
where we have suppressed all arguments in the coefficient functions ${\cal C}_{k, i}$ for brevity.
The derivative $\partial_\eta$ has been turned into the derivative
$\partial_z$ thanks to Eq.~(\ref{eq:eta-def}) and all partial derivatives 
in Eq.~(\ref{eq:cdiff}) have been made explicit.

The boundary term in Eq.~(\ref{eq:partial-F2c}) vanishes explicitly for the NLO scheme transformation in Eq.~(\ref{eq:F2c-mbar}), 
i.e. for $\partial_m F_k^{(0)}(m)$, 
because the Born contribution behaves for $\beta \to 0$ as $C_{2, g}^{(0)} \sim {\cal O}(\beta)$ and 
$C_{L, g}^{(0)} \sim {\cal O}(\beta^3)$.
Thus, $C_{k, g}^{(0)}$ vanishes in the last line of Eq.~(\ref{eq:partial-F2c}) if evaluated at $z=z^{\rm max}$
We note however, that the boundary term may be completely removed to all orders by means of partial integration with respect to the PDFs. 
With integration-by-parts in $z$ we find for Eq.~(\ref{eq:partial-F2c}),
\begin{eqnarray}
\label{eq:ibp-allorder-F2c}
  \partial_m F_k^{(l)}(m) &=&
  \sum\limits_{i = q,{\bar{q}},g} \,\,
  \int\limits_{x}^{z^{\rm max}}\,
  {dz \over z} \,\, f_{i}\left({x \over z}\right)\,
\left\{
  {2 \over m}\, z\, {\cal C}_{k, i}^{(l)}
- {2 \over m} \xi\, \partial_\xi {\cal C}_{k, i}^{(l)}
- {\muf \over m} \partial_\muf {\cal C}_{k, i}^{(l)}
\right\}
\\
& &
\nonumber
-
  \sum\limits_{i = q,{\bar{q}},g} \,\,
\int\limits_{x}^{z^{\rm max}}\,
  {dz \over z} \,\, 
  \left( z\, \partial_z\, f_{i}\left({x \over z}\right) \right)\, \left\{ 
      {2 \over m}\, (1-z)\, {\cal C}_{i,k}^{(l)}
    \right\}
    \, .
\end{eqnarray}

The NNLO scheme transformation in Eq.~(\ref{eq:F2c-mbar}) requires as the only
additional ingredient the second derivative $\partial_m^2$ for the Born term $F_k^{(0)}(m)$, 
i.e.
\begin{eqnarray}
\label{eq:ibp-second-diff}
  \partial_m^2 F_k^{(0)}(m) 
&=&
  - {2 \over m^2}\, \left\{ {m \over 2} \partial_m F_k^{(0)}(m) \right\}
  + {2 \over m}\, \partial_m\, \left\{ {m \over 2} \partial_m F_k^{(0)}(m) \right\}
\, ,
\end{eqnarray}
which can be quickly evaluated using Eq.~(\ref{eq:ibp-allorder-F2c}) and
computing the explicit derivative of the coefficient functions ${\cal C}_{k, g}^{(0)}$.
We note, that 
$\partial_\xi {\cal C}_{2, g}^{(0)} \sim {\cal O}(\beta)$ 
and 
$\partial_\xi {\cal C}_{L, g}^{(0)} \sim {\cal O}(\beta^3)$,  
so that they also vanish if evaluated at $z=z^{\rm max}$.

Thus, we are finally in a position to put everything together through NNLO and 
we arrive at the following explicit expression for Eq.~(\ref{eq:F2c-mbar}), 
\begin{eqnarray}
  \label{eq:final}
  F_k &=&
  \alphas\, F_k^{(0)}
  + 
  \alphas^2\, F_k^{(1)}
  + 
  \alphas^3\, F_k^{(2)}
\\
& &
  +
  \alphas^2\,
  \sum\limits_{i = q,{\bar{q}},g} \,\,
  \int\limits_{x}^{z^{\rm max}}\,
  {dz \over z} \,\, f_{i}\left({x \over z}\right)\
  \,\, 2\, d^{(1)}\, 
  \left\{ 
  {m \over 2}\,  \partial_m
    {\cal C}_{k, i}^{(0)}
  \right\}
\nonumber \\
& &
  +
  \alphas^3\, 
  \sum\limits_{i = q,{\bar{q}},g} \,\,
  \int\limits_{x}^{z^{\rm max}}\,
  {dz \over z} \,\, f_{i}\left({x \over z}\right)\
  \left(2\, d^{(2)} - \left(d^{(1)}\right)^2 \right)\,
  \left\{ 
  {m \over 2}\,  \partial_m
    {\cal C}_{k, i}^{(0)}
  \right\}
\nonumber \\
& &
  +
  \alphas^3\, 
  \sum\limits_{i = q,{\bar{q}},g} \,\,
  \int\limits_{x}^{z^{\rm max}}\,
  {dz \over z} \,\, 
  \left[
  f_{i}\left({x \over z}\right)\,
  \left\{
    z\, 
    - \xi\, \partial_\xi 
    - {\muf \over 2} \partial_{\muf}
  \right\}\, 
  +
  {x \over z} f^\prime_{i}\left({x \over z} \right) 
  (1-z) 
\right]\, 
  \,\, 2\, d^{(1)}\, 
{\cal C}_{i,k}^{(1)}
\nonumber \\
& &
  +
  \alphas^3\, 
  \sum\limits_{i = q,{\bar{q}},g} \,\,
  \int\limits_{x}^{z^{\rm max}}\,
  {dz \over z} \,\, 
\left[
  f_{i}\left({x \over z}\right)\,
  \left\{
    z\, 
    - \xi\, \partial_\xi 
  \right\}\, 
  +
  {x \over z} f^\prime_{i}\left({x \over z} \right) 
  (1-z) 
\right]\, 
2\, \left(d^{(1)}\right)^2\, 
  \left\{ 
  {m \over 2}\,  \partial_m
    {\cal C}_{k, i}^{(0)}
  \right\}
\, ,
\nonumber
\end{eqnarray}
where the renormalization scale has been fixed at $\mur = \mm$, 
i.e. $\alphas = \alphas(\mm)$ for the strong coupling constant.
The full dependence on $\mur$ can be constructed using the renormalization group equation. 
With the standard expression for the running coupling 
(and the coefficients of the QCD beta function),
it is easy to restore the complete renormalization scale dependence of $\alphas$ in Eq.~(\ref{eq:final}),
\begin{equation}
\label{eq:asrun}
  \alphas(\mm) \,=\, 
  \alphas(\mur) \left( 
    1 + \alphas(\mur) L_R \* \beta_0 + \alphas(\mur)^2 ( \beta_1 L_R + \beta_0^2 L_R^2)\right)
  \, ,
\end{equation}
where we have abbreviated $L_R = \ln(\mur^2 / \mm^2)$.

As explained above, Eq.~(\ref{eq:final}) is exact to NLO.
At the NNLO level, ${\cal C}_{k, i}^{(2)}$ is currently unknown. 
Our approximation NNLO$_{\rm approx}$ in Eq.~(\ref{eq:final}) uses the threshold enhanced terms
of Ref.~\cite{Presti:2010pd} to estimate the dominant corrections at two loops. 
All explicit dependence on the renormalization and factorization scales and
the terms accounting for the scheme transformation from the pole to the
running mass at NNLO are, however, exact in Eq.~(\ref{eq:final}).

\bigskip

The CC case is conceptually much simpler. 
Moreover, in our analysis, we confine ourselves to the NLO case only 
so that the relevant formulae is much shorter. 
One commonly defines structure functions ${\cal F}_k$ 
which are related to the $F_k$'s via the following relations:
\begin{eqnarray}
\label{eq:ccredsf1}
F_1 \,=& {\cal F}_1 &=\, a_1\, {\cal F}_1 
\, ,\\
\label{eq:ccredsf2}
F_2 \,=& 2 \chi\, {\cal F}_2 &=\, a_2\, {\cal F}_2 
\, ,\\
\label{eq:ccredsf3}
F_3 \,=& 2\, {\cal F}_3 &=\, a_3\, {\cal F}_3 
\, ,
\end{eqnarray}
i.e. $a_1=1$, $a_2=2\, \chi$ and $a_3=2$ and where we have introduced the quantity
\begin{equation}
\label{eq:chi}
\chi={x \over\lambda} 
\, ,
\end{equation}
so that Bjorken $x$ varies in the range $ 0 < x \leq \lambda $ and 
\begin{equation}
\label{eq:lamb}
\lambda \,= \, {1 \over {1 + {m^2\over Q^2}}} \,= \, {\xi \over {1 + \xi}}
\, .
\end{equation}

In the CC case, the coefficient functions corresponding to Eq.~(\ref{eq:totalF2c}) 
can be expanded to NLO in the strong coupling with $\alphas = \alphas(\mur)$,
\begin{equation}
  \label{eq:partonCcc-exp}
  {\cal C}_{k, i}(z,\xi,\murs,\mufs) =
  {\cal C}_{k, i}^{(0)} +  \alphas\, {\cal C}_{k, i}^{(1)}
  \, ,
\end{equation}
where ${\cal C}_{k, q}^{(0)} \simeq \delta(1-z)$ (up to the CKM parameters) 
and ${\cal C}_{k, g}^{(0)} = 0$ for $k=1,2,3$ due to Eq.~(\ref{eq:ccborn}).
The expressions for ${\cal C}_{k, i}^{(1)}$ are all given in Refs.~\cite{Gottschalk:1980rv,Gluck:1996ve}.
As in the NC case, the conversion to the \MSbar mass starts off from a perturbative expansion as in Eq.~(\ref{eq:F2c-mpole}),
which we can convert to the \MSbar mass as in Eq.~(\ref{eq:F2c-mbar}).
Restricting ourselves to NLO, we relate
\begin{equation}
  \label{eq:F2cc-mpole}
  F_k(m) =
  F_k^{(0)}(m) +  \alphas\, F_k^{(1)}(m) 
  \, ,
\end{equation}
to
\begin{eqnarray}
  \label{eq:F2cc-mbar}
  F_k(\mbar) &=&
  F_k^{(0)}(\mbar) 
  + 
  \alphas\, \left( 
    F_k^{(1)}(\mbar) 
    + \mbar\, d^{(1)} \partial_m F_k^{(0)}(m) \biggr|_{m=\mbar}
  \right) 
  \, ,
\end{eqnarray}
where matching is done again at the scale $\mur = \mbar$ and we just have to provide the explicit expression 
\begin{eqnarray}
  \label{eq:F2cc-mbar-v2}
  \mbar\, d^{(1)} \partial_m a_k F_k^{(0)}(m) \biggr|_{m=\mbar} 
&=&
  2\, (\chi - x)\, d^{(1)} 
  \sum\limits_{i = q_1} \,\,
  \left| V_{i\, q_2} \right|^2
  \left( \partial_\chi\, a_k\, 
  f_{i}\left(\chi\right)
  \right) \biggr|_{m=\mbar} 
  \, ,
\end{eqnarray}
where the sum ranges over all light flavors $q_1$, while $q_2$ is heavy.
$V_{q_1\, q_2}$ denotes the respective CKM matrix element 
and the coefficients $a_k$ are defined in Eqs.~(\ref{eq:ccredsf1})--(\ref{eq:ccredsf3}). 
The partial derivative from $ \partial_m \chi$ has again been made explicit.

\bigskip

This completes our discussion on the explicit conversion of the 
heavy-quark DIS structure functions from the pole to the \MSbar mass.
Note, that we have confined ourselves entirely to the 
so-called fixed-flavor-number scheme (FFNS),  i.e. to a situation, where we work 
with a fixed number $n_f$ of light-quark flavors.
This is an absolutely adequate approach e.g. for the analysis of existing DIS data.
In contrast, there exist variable flavor number schemes (VFNS) which relate 
the DIS structure functions for $n_f$ light flavors to those for $n_f+1$ light flavors 
(see~\cite{Alekhin:2009ni,Forte:2010ta} for an extensive discussion). 
In a VFNS the necessary matching involves certain massive operator matrix elements
which, for consistency also need to be evaluated in the renormalization scheme with a running mass. 
To NNLO, all relevant formulae has been given in Ref.~\cite{Bierenbaum:2009mv}.
Also, in practice, one often relates the matching scale of the strong coupling to
the heavy-quark mass, i.e. $\alphas^{(n_f)}(\kappa m) \to \alphas^{(n_f+1)}(\kappa m)$ with some constant $\kappa$. 
Then the necessary decoupling coefficients depend beyond NLO 
on the chosen mass renormalization (see e.g. \cite{Chetyrkin:2000yt}).
In summary, the implementation of a running mass for heavy-quark DIS in a VFNS 
can be performed in a straight forward manner and poses no further problems.

\section{Results}
\label{sec:res}

We are now in a position to look at the phenomenological implications of the \MSbar mass and 
also to discuss the role of the heavy-quark mass parameter in PDF fits.
Despite of its short-comings, all current global fits of PDFs employ the pole mass scheme for the heavy quarks.
To illustrate this point, we summarize in Tab.~\ref{tab:polemassvalues} the values taken 
by the six groups which are currently active in global fits PDF:  
ABKM~\cite{Alekhin:2009ni}, 
HERAPDF~\cite{herapdf:2009wt}, 
GJR~\cite{Gluck:2007ck},
MSTW~\cite{Martin:2009iq},
CTEQ~\cite{Nadolsky:2008zw} and  
NNPDF~\cite{Ball:2008by}.
All numerical values, especially those for the charm quark mass, are 
systematically lower than the pole masses obtained from the particle data group (PDG) values 
for the world average~\cite{Nakamura:2010pdg}. 
Simple kinematical considerations show, that a smaller charm quark mass can potentially compensate 
large missing higher order perturbative corrections. 
The latter have been shown to be sizable at NLO and even at NNLO$_{\rm approx}$ when using a pole mass.
Evidently, global fits of PDFs which incorporate heavy-quark DIS data are very
sensitive to the theory treatment of heavy quarks and 
depending on the chosen mass parameter (and scheme, see~\cite{Alekhin:2009ni} for a discussion of FFNS and VFNS) 
the resulting differences in the PDFs can be sizable.
\begin{table}[t!]
\centering
{\small
\begin{tabular}{c|c|c|c|c|c|c|c}
   [GeV] 
  & PDG
  & ABKM
  & GJR
  & HERAPDF
  & MSTW
  & CTEQ
  & NNPDF
\\[1ex]
\hline  & & & & & & & \\[-2ex]
$m_c$ 
  & 1.66\,\,$^{+0.09}_{-0.15}$ 
  & 1.5\,\,$^{+0.25}_{-0.25}$
  & 1.3
  & 1.4
  & 1.4 
  & 1.3
  & 
$\sqrt{2}$
\\ [1ex]
$m_b$ 
  & 4.79\,\,$^{+0.19}_{-0.08}$ 
  & 4.5\,\,$^{+0.5}_{-0.5}$
  & 4.2
  & 4.75
  & 4.75 
  & 4.5 
  & 4.3
\end{tabular}
}
\caption{\small
\label{tab:polemassvalues}
 The pole mass values taken as input in recent global fits of PDFs.
 The quoted PDG values are obtained from the 
 \MSbar values in Eqs.~(\ref{eq:mcpdg}), (\ref{eq:mbpdg}) 
 using the two-loop conversion in Eq.~(\ref{eq:mpole-mbar}). 
}
\end{table}

\bigskip

\begin{figure}[th!]
\centering
    {
      \includegraphics[scale = 0.675]{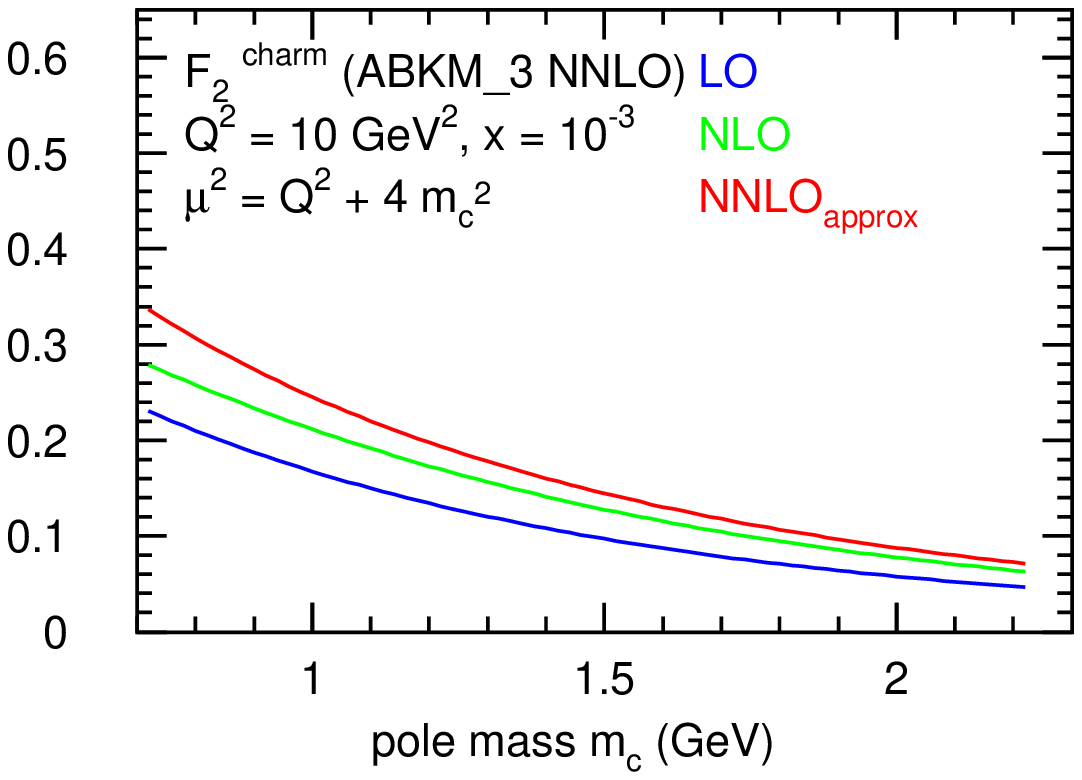}
    }
\hspace*{10mm}
    {
      \includegraphics[scale = 0.675]{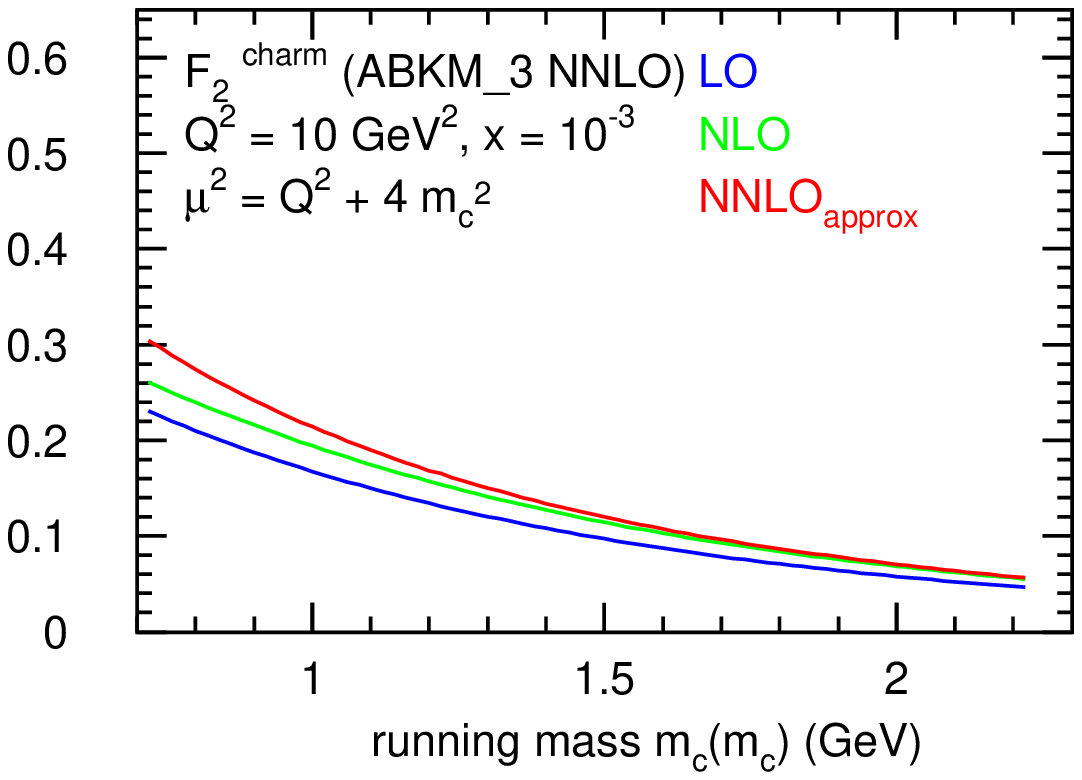}
    }
\vspace*{-1mm}
\caption{\small
  \label{fig:msbarscale-cen}
 The mass dependence of the NC charm structure function $F_2^p$ for a proton 
 with $Q^2 = 10 \GeV^2$, $x=10^{-3}$ 
 and $\mur = \muf = \sqrt{Q^2 + 4 m_c^2}$ using the PDFs of~\cite{Alekhin:2009ni}.
 The charm-quark mass is taken in the on-shell scheme (left) and
 in the \MSbar scheme (right) at LO (blue), NLO (green) and NNLO$_{\rm approx}$ (red).
}
\centering
\vspace*{10mm}
    {
      \includegraphics[scale = 0.675]{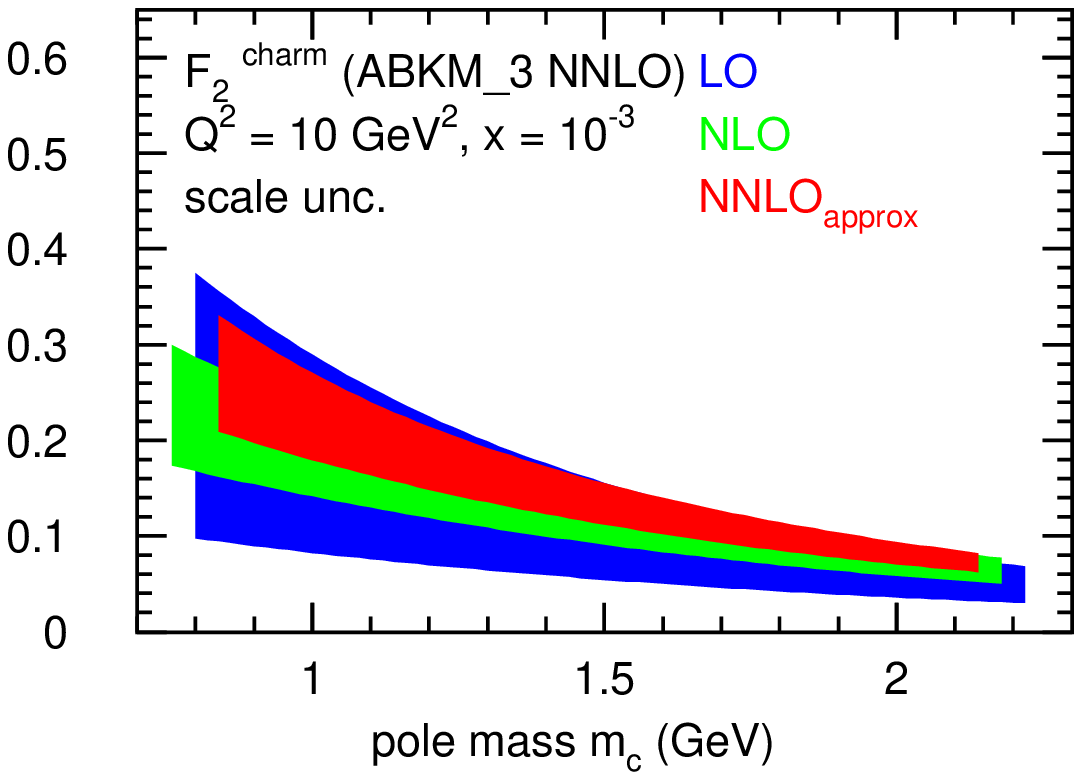}
    }
\hspace*{10mm}
    {
      \includegraphics[scale = 0.675]{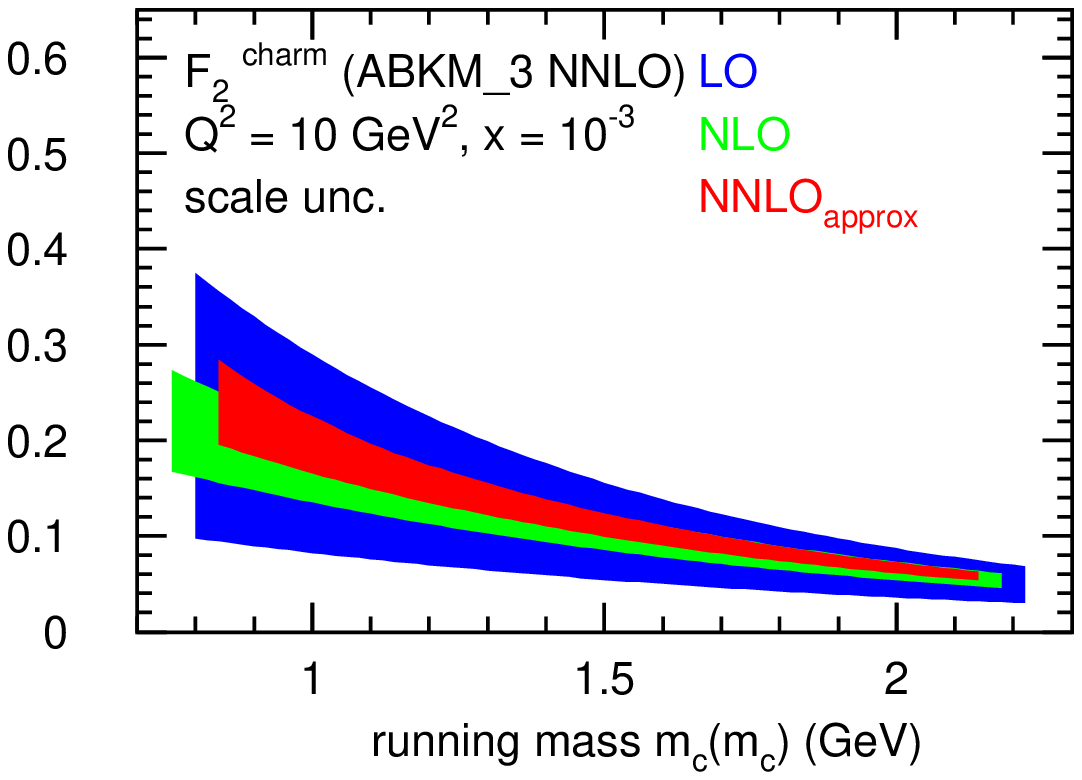}
    }
\vspace*{-1mm}
\caption{\small
  \label{fig:msbarscale}
 Same as in Fig.\ref{fig:msbarscale-cen}. The band denotes the independent
 variation of the scales 
 $\mur,\muf = \kappa \sqrt{Q^2 + 4 m_c^2}$ in the range $\kappa \in [1/2, 2]$.
}
\centering
\vspace*{10mm}
    {
      \includegraphics[scale = 0.65]{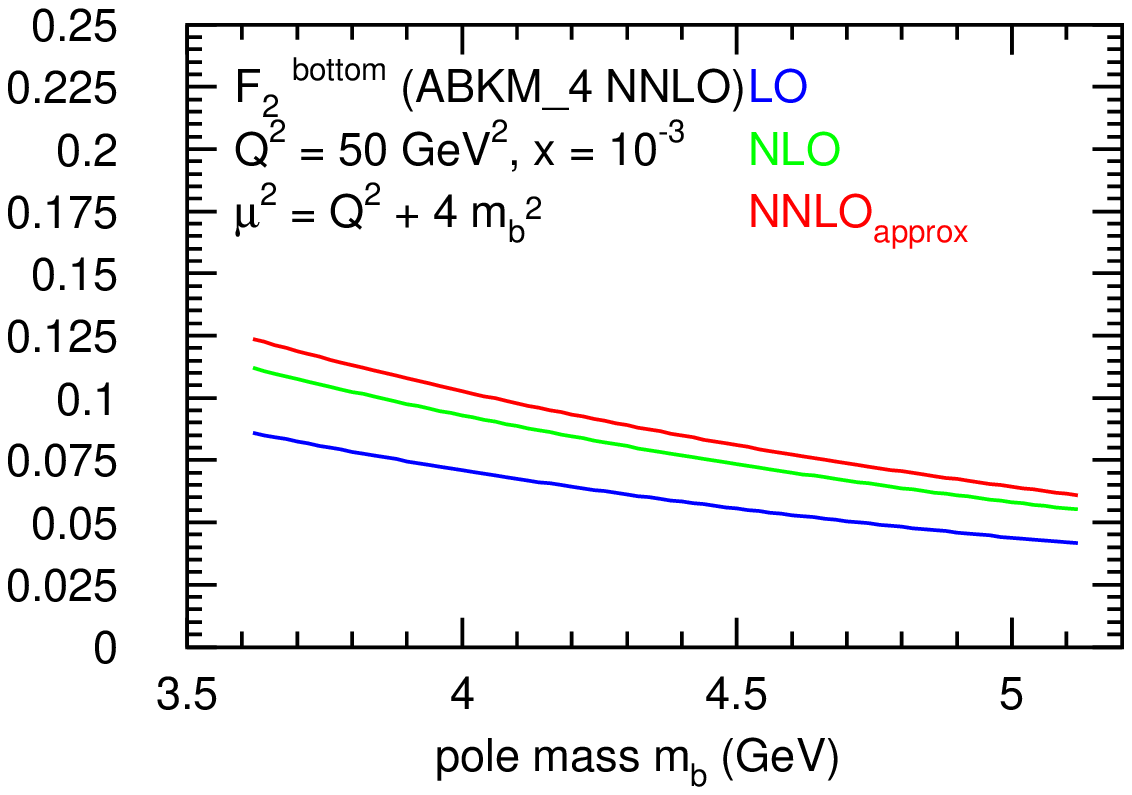}
    }
\hspace*{10mm}
    {
      \includegraphics[scale = 0.65]{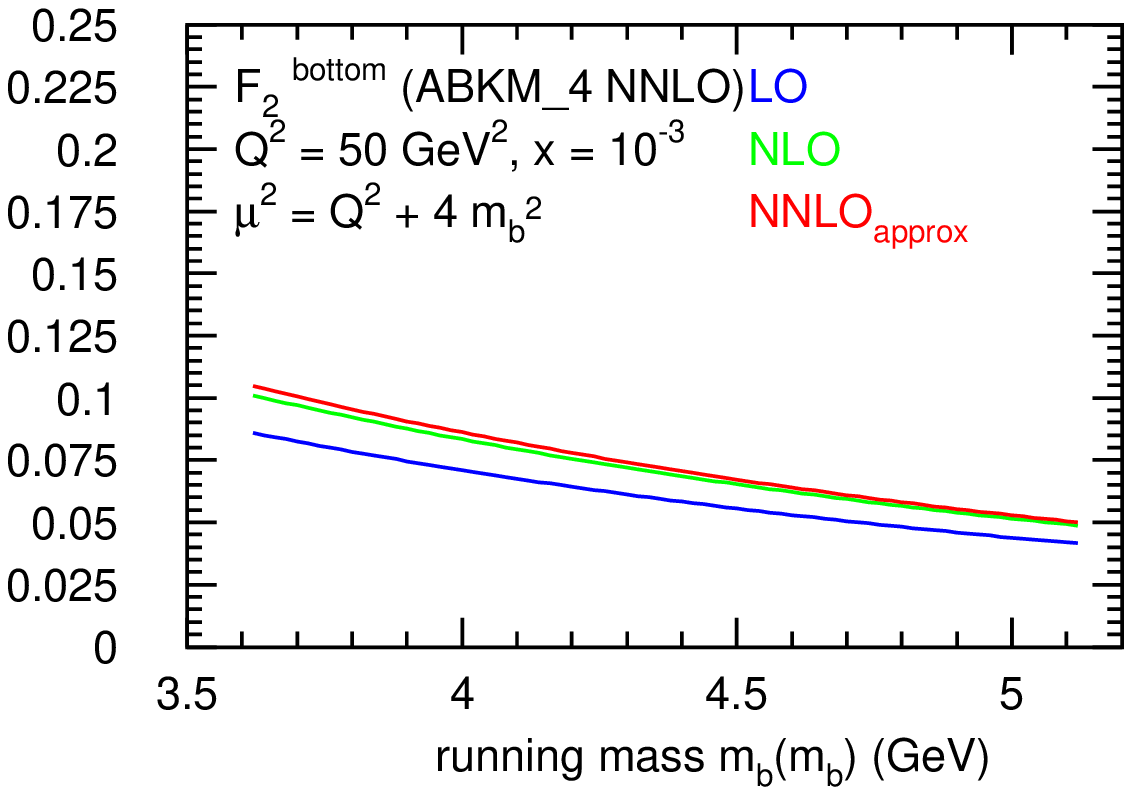}
    }
\vspace*{-1mm}
\caption{\small
  \label{fig:msbarscale-bottom-cen}
 The mass dependence of the NC bottom structure function $F_2^p$ for a proton 
 with $Q^2 = 50 \GeV^2$, $x=10^{-3}$ 
 and $\mur = \muf = \sqrt{Q^2 + 4 m_c^2}$.
 The bottom-quark mass is taken in the on-shell scheme (left) and
 in the \MSbar scheme (right) at LO (blue), NLO (green) and NNLO$_{\rm approx}$ (red).
}
\end{figure}

In Figs.~\ref{fig:msbarscale-cen} and~\ref{fig:msbarscale} we study first the
case of charm quark electro-production, i.e. electron-proton scattering.
We plot the NC charm structure function $F_2^p$ at LO, NLO and NNLO$_{\rm approx}$ 
using the 3-flavor PDF set of Ref.~\cite{Alekhin:2009ni}.
We have been careful to restrict the kinematics in $x$ and $Q^2$ 
(in Figs.~\ref{fig:msbarscale-cen}, \ref{fig:msbarscale} $Q^2 = 10 \GeV^2$, $x=10^{-3}$) 
to the region, where our threshold approximation underlying the NNLO$_{\rm approx}$ prediction 
is under control, see~\cite{Laenen:1998kp,Alekhin:2008hc}. 

Comparing the central values of the predictions for $F_2^p$ as a function of the pole mass 
in Fig.~\ref{fig:msbarscale-cen} (left) with those for an \MSbar mass $m_c(m_c)$ 
in Fig.~\ref{fig:msbarscale-cen} (right), 
we observe a much improved convergence of the perturbative expansion in the latter case.
Already at NLO the size of the QCD corrections is much reduced for an \MSbar mass.
E.g. for a pole mass of $m_c = 1.5$~GeV we find for $F_2^p$ a relative increase of $32\%$ at NLO over the LO prediction 
and another $13\%$ for the NNLO$_{\rm approx}$ prediction normalized to the NLO one.
This is to be compared with the numbers for the \MSbar mass. 
At $m_c(m_c) = 1.3$~GeV we find relative corrections of $17\%$ at NLO and only $6\%$ at NNLO$_{\rm approx}$.
Likewise, as we vary the renormalization and the factorization scale $\mur$
and $\muf$ independently by a factor of two around a central value 
chosen to be $\mur^2 = \muf^2 = Q^2 + 4 m_c^2$ in Figs.~\ref{fig:msbarscale-cen} and \ref{fig:msbarscale}  
we note a substantial reduction in the spread to the predictions and a greatly reduced theoretical uncertainty 
when using the running mass.
This scale variation is illustrated by the respective bands in Fig.~\ref{fig:msbarscale}.

\begin{figure}[th!]
\centering
    {
      \includegraphics[scale = 0.65]{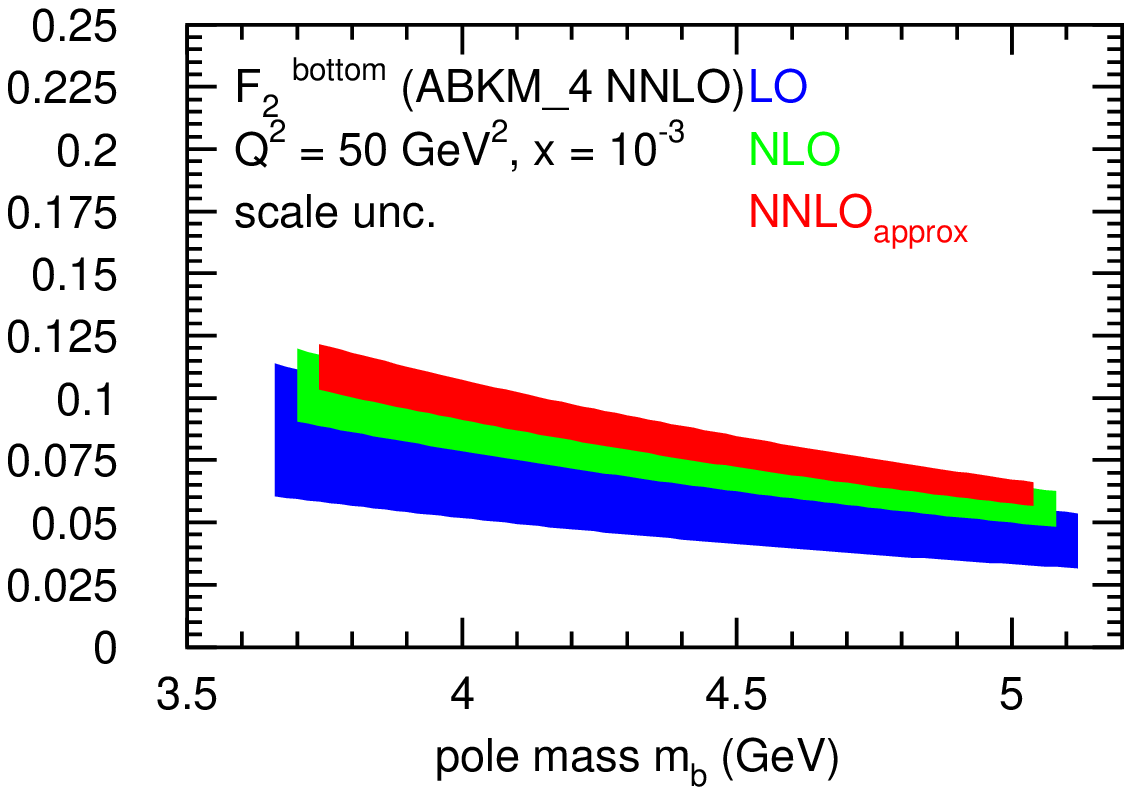}
    }
\hspace*{10mm}
    {
      \includegraphics[scale = 0.65]{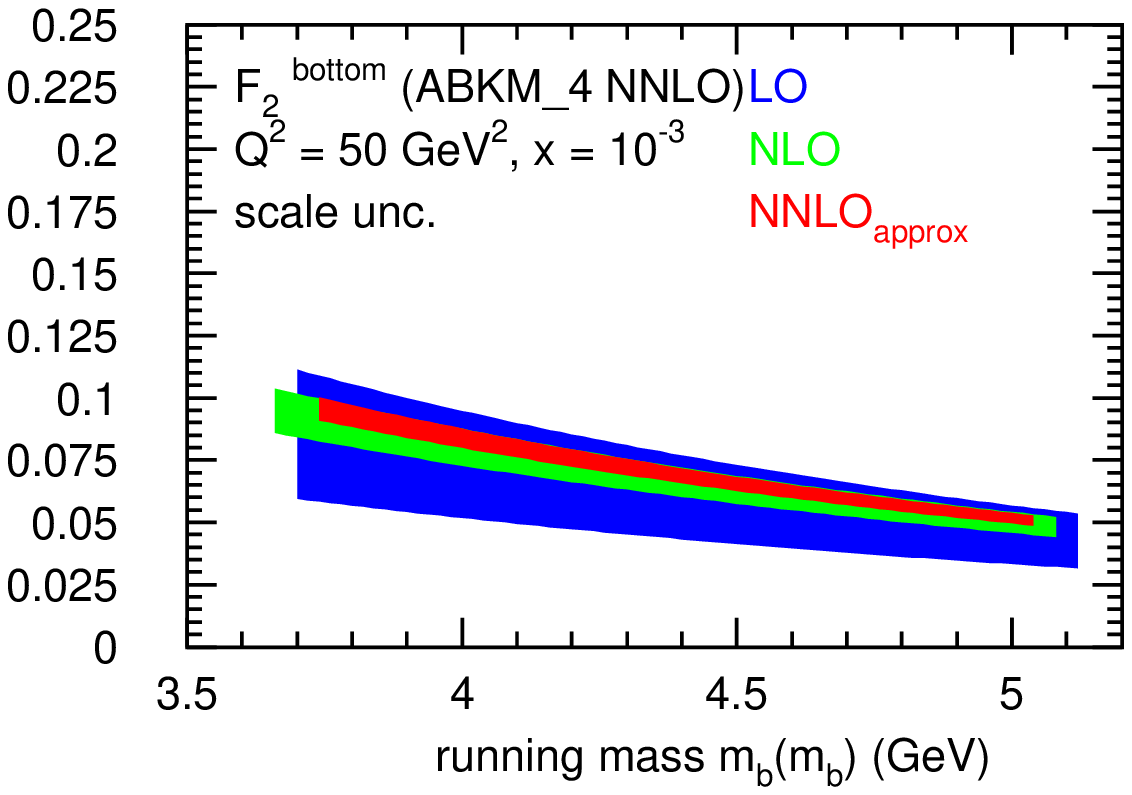}
    }
\vspace*{-1mm}
\caption{\small
  \label{fig:msbarscale-bottom}
 Same as in Fig.\ref{fig:msbarscale-bottom-cen}. The band denotes the independent
 variation of the scales 
 $\mur,\muf = \kappa \sqrt{Q^2 + 4 m_c^2}$ in the range $\kappa \in [1/2, 2]$.
}
\centering
\vspace*{10mm}
    {
      \includegraphics[scale = 0.675]{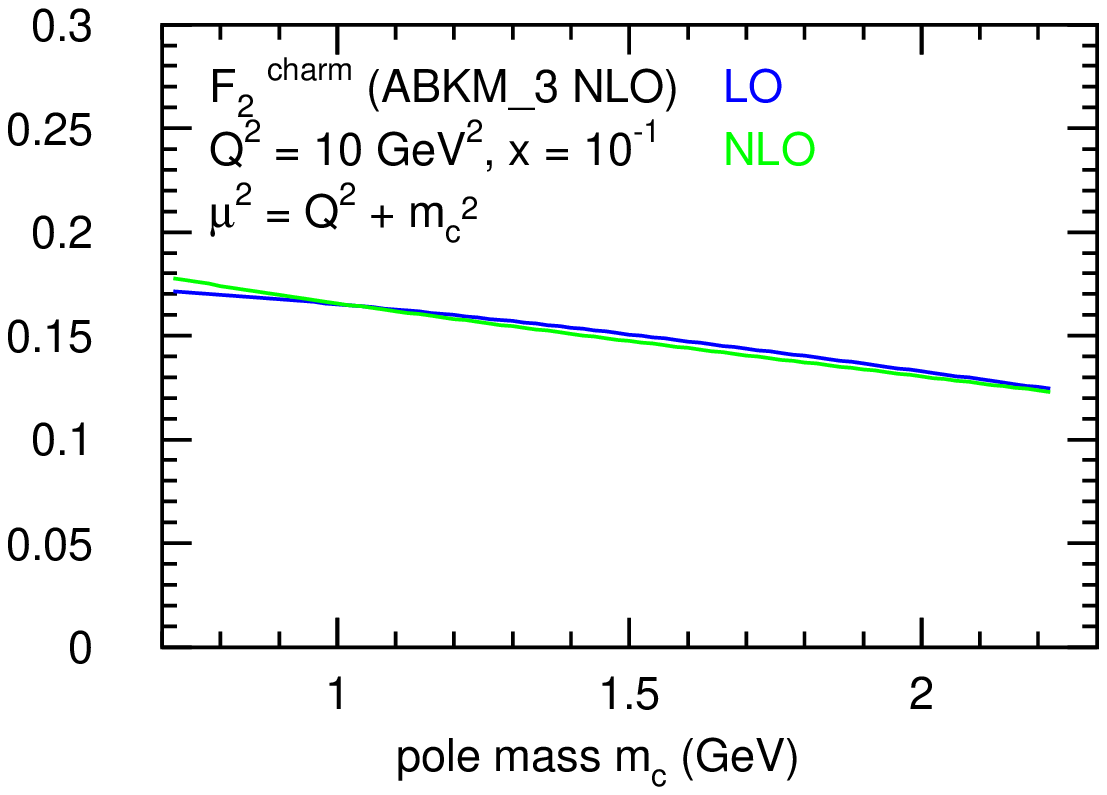}
    }
\hspace*{10mm}
    {
      \includegraphics[scale = 0.675]{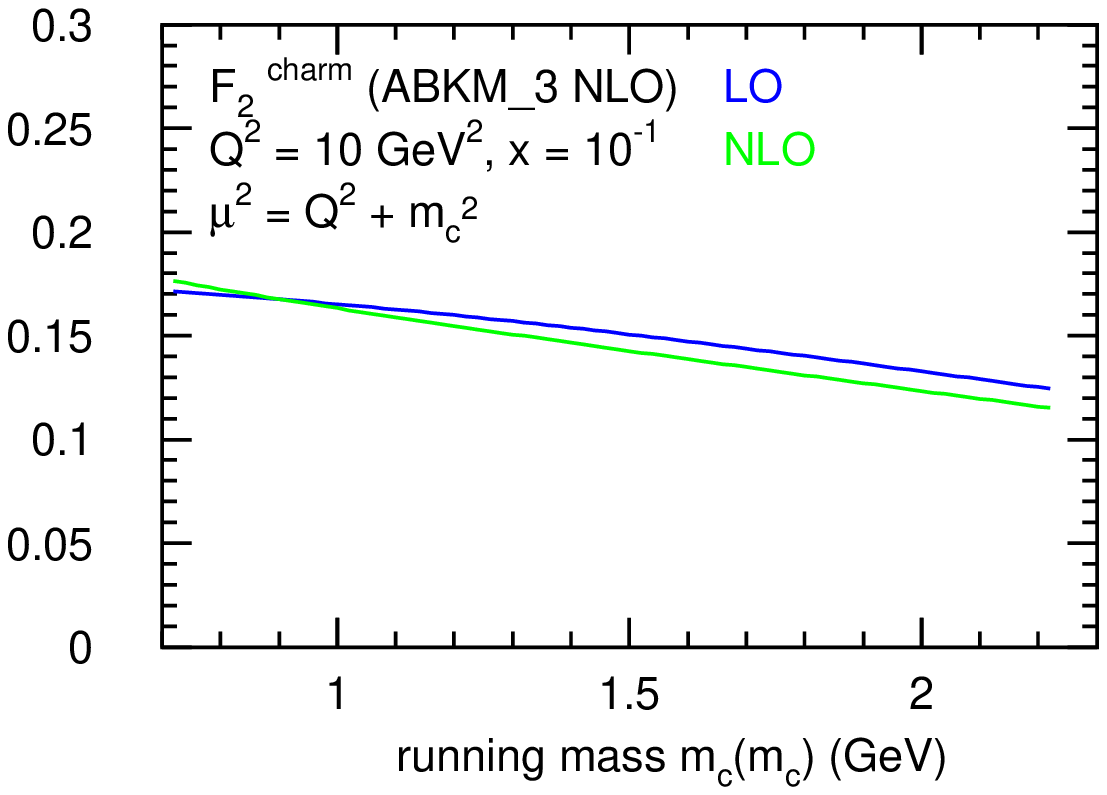}
    }
\vspace*{-1mm}
\caption{\small
  \label{fig:CC-msbarscale-cen}
 The mass dependence of the CC charm structure function $F_2^N$ for a nucleon 
 with $Q^2 = 10 \GeV^2$, $x=10^{-1}$ 
 and $\mur = \muf = \sqrt{Q^2 + m_c^2}$ using the PDFs of~\cite{Alekhin:2009ni}.
 The charm-quark mass is taken in the on-shell scheme (left) and
 in the \MSbar scheme (right) at LO (blue) and NLO (green).
}
\centering
\vspace*{10mm}
    {
      \includegraphics[scale = 0.675]{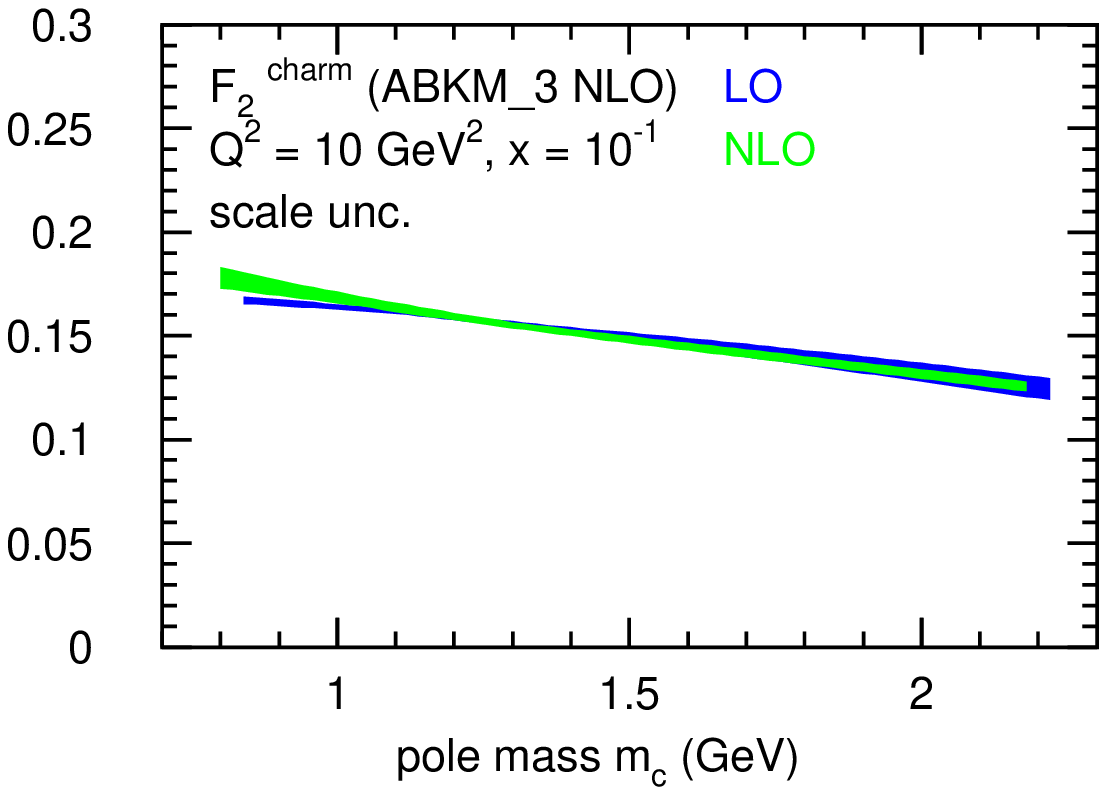}
    }
\hspace*{10mm}
    {
      \includegraphics[scale = 0.675]{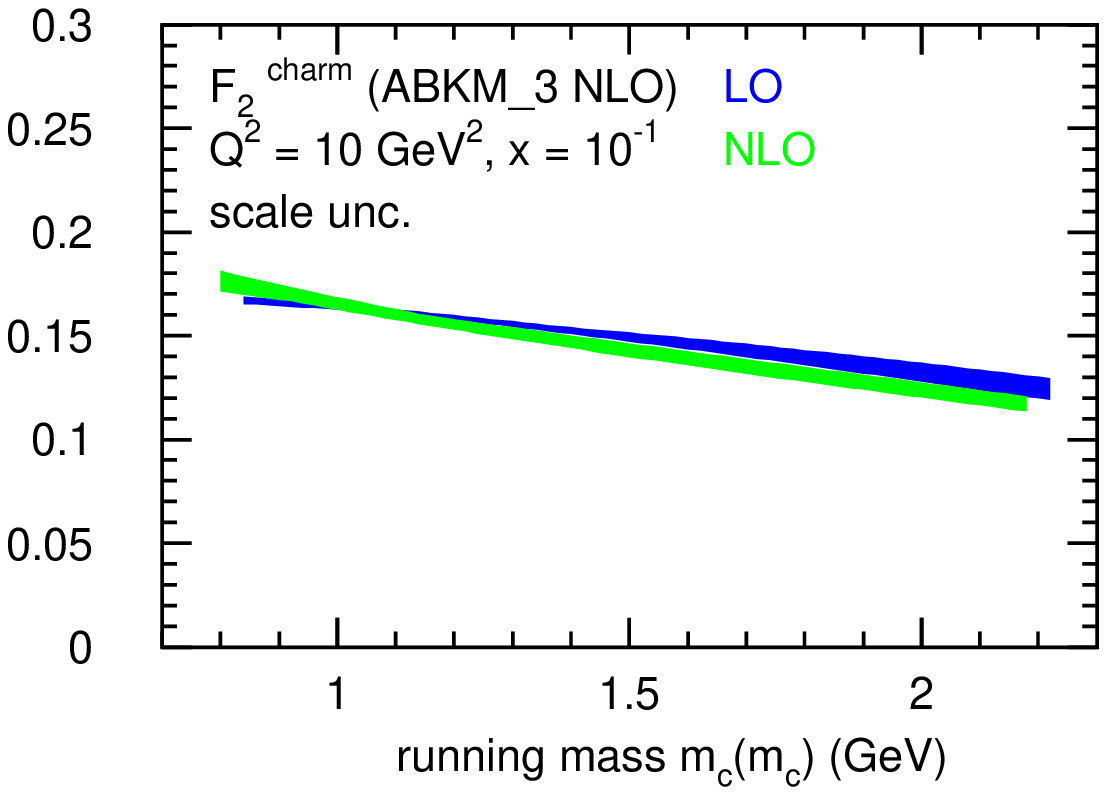}
    }
\vspace*{-1mm}
\caption{\small
  \label{fig:CC-msbarscale}
 Same as in Fig.\ref{fig:CC-msbarscale-cen}. The band denotes the independent
 variation of the scales 
 $\mur,\muf = \kappa \sqrt{Q^2 + m_c^2}$ in the range $\kappa \in [1/2, 2]$.
}
\end{figure}

Similar observations hold also for the NC DIS production of bottom quarks, 
see Figs.~\ref{fig:msbarscale-bottom-cen} and~\ref{fig:msbarscale-bottom}.
The predictions at $Q^2 = 50 \GeV^2$ and $x=10^{-3}$ have been obtained with 
the 4-flavor PDF set of~\cite{Alekhin:2009ni}. 
Again, in the case of the \MSbar mass the apparent convergence is much improved and, 
in particular, the relative size of NNLO$_{\rm approx}$ corrections is a few
per cent only over the whole mass range considered for $m_b(m_b)$.

In Figs.~\ref{fig:CC-msbarscale-cen} and \ref{fig:CC-msbarscale} we investigate charm quark production in 
neutrino-nucleon DIS assuming an isoscalar target.
We plot the CC charm structure function $F_2^N$ for a nucleon at LO and NLO for $Q^2 = 10 \GeV^2$, $x=10^{-1}$ which corresponds 
to the typical kinematics of fixed-target neutrino-nucleon experiments.
In comparison, the impact of higher order perturbative corrections is less than 
in the NC case discussed before.

We compare the central predictions using a pole mass (Fig.~\ref{fig:CC-msbarscale-cen} left) 
with those that employ the running mass (Fig.~\ref{fig:CC-msbarscale-cen} right) 
and the observed differences are rather marginal.
This can easily be understood because the conversion to the \MSbar mass only involves 
(e.g. for a $W^+$-boson on isoscalar nucleon target) 
the derivatives of the up, down and the strange quark PDFs, which are all numerically rather small.
Again, we also vary the renormalization and the factorization scale $\mur$
and $\muf$ independently by a factor of two around the central value $\mur^2 = \muf^2 = Q^2 + m_c^2$. 
The comparison is shown in Fig.~\ref{fig:CC-msbarscale} and the are no significant changes.
In summary, we observe that the impact of the scheme change from a pole to a running mass 
is much more pronounced in the case of NC DIS than in the case of CC DIS.

\bigskip

We have demonstrated a clear improvement of the theoretical predictions for
DIS heavy-quark structure functions by using the running mass.
An immediate application of our results consists therefore in the direct determination 
of heavy-quark \MSbar masses from the available DIS data.
For reference in the following, we list the values for the charm and bottom masses in the \MSbar scheme as quoted in 
the 2010 edition of the PDG~\cite{Nakamura:2010pdg},
\begin{eqnarray}
  \label{eq:mcpdg}
  m_c(m_c) &=& 1.27\,\,^{+0.07}_{-0.09}\,\, {\rm GeV} 
\, ,
\\
  \label{eq:mbpdg}
  m_b(m_b) &=& 4.19\,\,^{+0.18}_{-0.06}\,\, {\rm GeV} 
\, .
\end{eqnarray}

To start with, we can use the manifest dependence of the structure functions $F_k$ on the heavy-quark mass 
to estimate the prospects of this approach.
The relative uncertainty of such a mass determination is related to the
corresponding uncertainty on the measurements of $F_k$ as follows.
Neglecting non-linear terms, a fit to the central prediction e.g. for $F_2^p$
in NC DIS (see Fig.~\ref{fig:msbarscale-cen}) yields, 
\begin{eqnarray}
  \label{eq:mass-sensitivity}
  {\Delta m_c \over m_c} &\simeq& 0.75 \, {\Delta F_2^p \over F_2^p}
  \, .
\end{eqnarray}
This implies that a measurement of the proton structure function $F_2^p$ 
with an accuracy of 10\% translates into a $0.75 \times 10\% = 7.5\%$ uncertainty 
of the charm-quark mass.
Thus, given the accuracy of current collider data from HERA (especially from HERA-II) 
an error on $m_c(m_c)$ of ${\cal O}({\rm few}) \%$ seems to be the ultimate
precision one can aim at in this approach.

\bigskip

For a quantitative comparison we conduct a phenomenological study similar to~\cite{Alekhin:2008hc,Alekhin:2008mb}, 
i.e. we perform a global fit of fixed-target (CCFR~\cite{Bazarko:1994tt}, NuTeV~\cite{Goncharov:2001qe})
and collider data~\cite{Chekanov:2003rb,Aaron:2009ut} in the FFNS (with $n_f = 3$) as a variant of ABKM~\cite{Alekhin:2009ni}.
In the analysis we have taken the same 25 parameters as in~\cite{Alekhin:2009ni} 
which include also the strong coupling $\alphas$ and the masses $m_c$ and $m_b$ 
besides the usual PDF parameters.
Interestingly, our fit does not return any sensitivity to the value of $m_b$. 
Therefore we have constrained the bottom mass $m_b(m_b)$ to its PDG value, i.e. Eq.~(\ref{eq:mbpdg}).
For the running mass of the charm-quark, however, our analysis displays very
good sensitivity and yields (depending on the order of perturbation theory) the following values 
\begin{eqnarray}
  \label{eq:mbarcnlo}
  m_c(m_c) &=& 1.26\,\, \pm 0.09 ({\rm exp})\,\, \pm 0.11 ({\rm th})\,\, 
  {\rm GeV}\qquad\qquad
  {\rm at\,\, NLO}
  \, ,
  \\[2ex]
  \label{eq:mbarcnnlo}
  m_c(m_c) &=& 1.01\,\, \pm 0.09 ({\rm exp})\,\, \pm 0.03 ({\rm th})\,\,
  {\rm GeV}\qquad\qquad
  {\rm at\,\, NNLO}_{\rm approx}
  \, ,
\end{eqnarray}
where the renormalization scale has been chosen $\mur = m_c$.
We consider our mass determination at NNLO$_{\rm approx}$ accuracy as our best estimate. 
Eq.~(\ref{eq:mbarcnnlo}) is the central result of this study 
and our determination is consistent with the world average 
at the level of $\pm 1.5\sigma$ for the quoted uncertainties.
Our NNLO$_{\rm approx}$ predictions are, of course, reliable only in a restricted kinematical range.
However, given that they are generally rather small, we consider the agreement
between the determinations at NLO and NNLO$_{\rm approx}$ also 
a very good indication on the stability of the perturbative description.

In Eqs.~(\ref{eq:mbarcnlo}) and (\ref{eq:mbarcnnlo}) 
the experimental and theoretical uncertainties on $m_c(m_c)$ have been quoted separately.
The former one is computed from the propagation of the statistical and systematic errors in the data, 
taking into account error correlations whenever available.
The theoretical uncertainty is estimated from the sensitivity due to variations 
of the renormalization and factorization scales $\mur$ and $\muf$ as follows.
All current global PDF determinations assume $\mur = \muf = Q$ in fits to DIS data  
(see e.g., the discussion in~\cite{Martin:2009iq}). 
This is the appropriate scale choice for massless structure functions and, generally, 
for large values of $Q$ when mass effects are negligible.
In order to retain sensitivity to mass effects, especially in the region of low $Q$, 
we therefore determine the variation of $F_2$ for the scale choice 
$\mur^2 = \muf^2 = Q^2 + \kappa m_c^2$ in the range $\kappa \in [0,8]$.
In this way, the variation of $F_2$ at NLO (NNLO$_{\rm approx}$) results
in the quoted uncertainty $\Delta m_c(m_c) = \pm 0.11$~GeV ($\Delta m_c(m_c) = \pm 0.03$~GeV).
For consistency, we have also checked, that the statistical quality of our fit is not deteriorated, 
if we use these different scale choices, i.e. the obtained value for $\chi^2$ changes by a few units only.

\begin{figure}[th!]
\centering
    {
    \includegraphics[width=16.0cm]{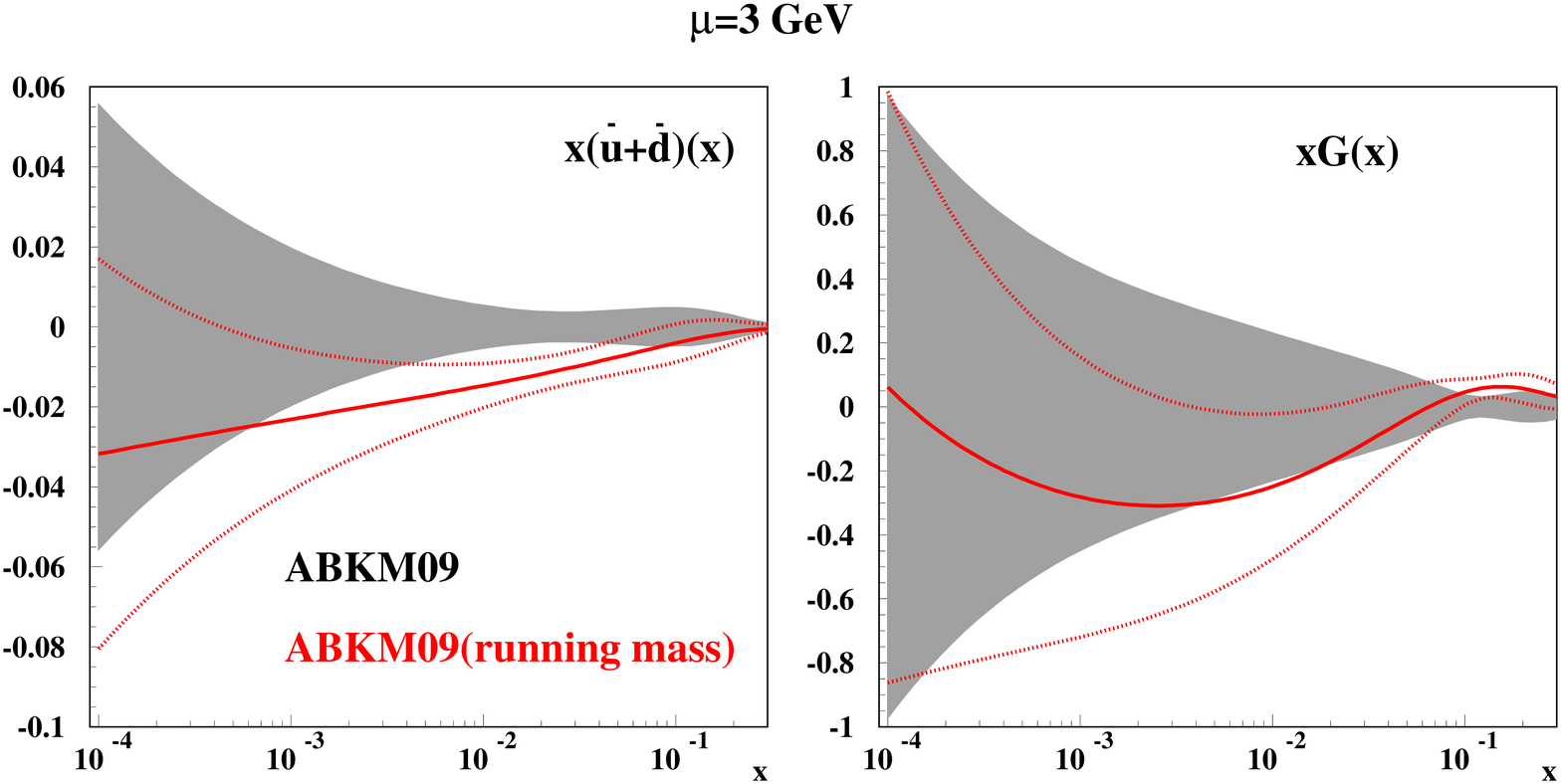}
    }
\vspace*{-1mm}
    \caption{ \small
      \label{fig:lq-pdfs}
      The light-quark (left) and gluon (right) PDFs obtained in the global fit:
      The dotted (red) lines denote the $\pm 1 \sigma$ band of absolute uncertainties  
      resulting from the fit of this paper and the solid (red) line indicates the
      central prediction with with running masses 1.18~GeV (charm) and 4.19~GeV (bottom).
      For comparison the shaded (grey) area represents the results of ABKM~\cite{Alekhin:2009ni}.
    }
\centering
\vspace*{10mm}
    {
    \includegraphics[width=14.25cm]{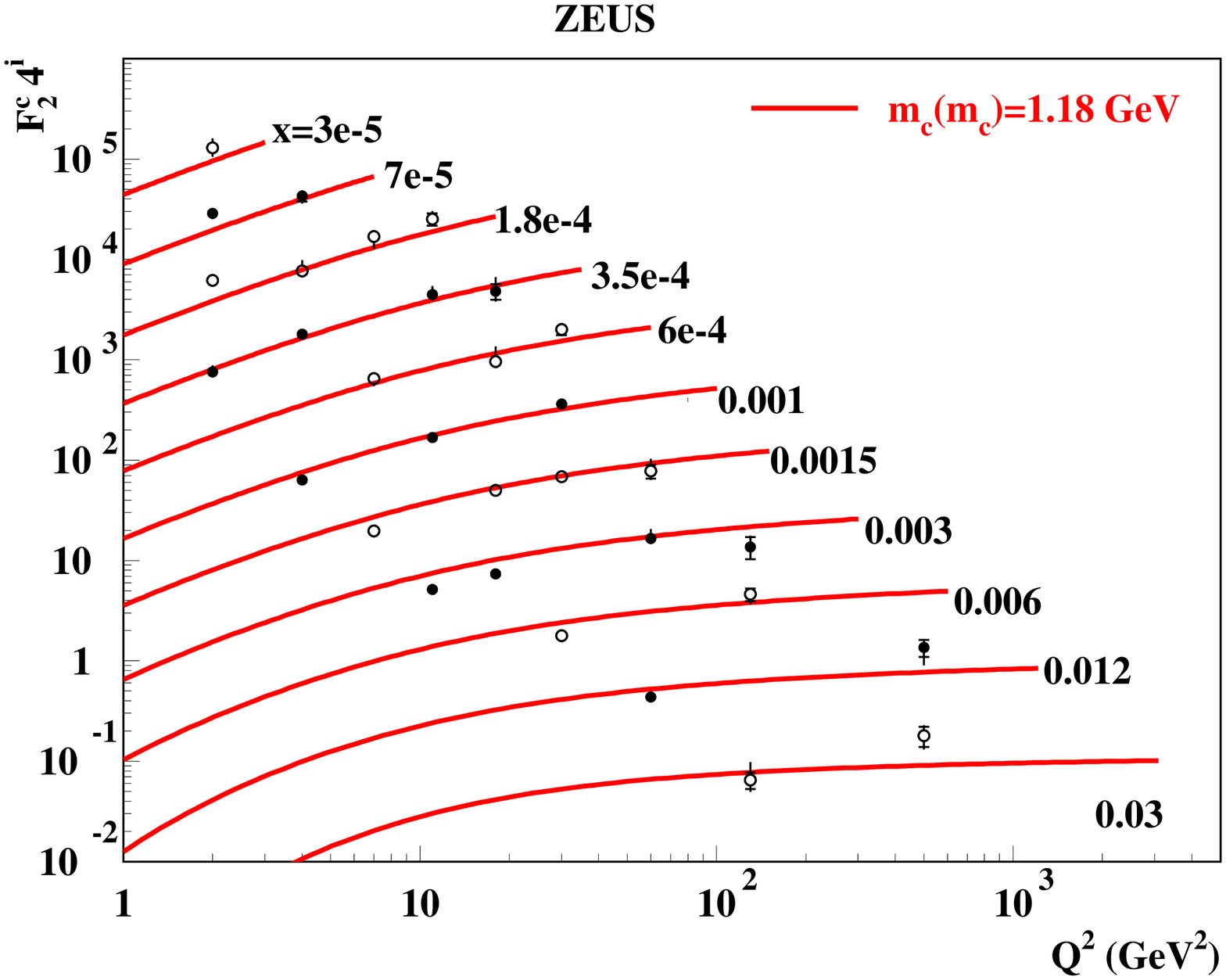}
    }
\vspace*{-6mm}
    \caption{ \small
      \label{fig:zeus}
      The predictions of the global fit for the charm structure function $F_2^c$ 
      compared to data of Ref.~\cite{Chekanov:2003rb} on $F_2^c$ 
      from Run I of HERA.
      The running mass $m_c(m_c) = 1.18$~GeV has been obtained in the variant
      of the fit with the PDG value Eq.~(\ref{eq:mcpdg}) as an additional constraint.
    }
\end{figure}

In comparison to Refs.~\cite{Alekhin:2009ni,Alekhin:2010iu} the shapes of the PDFs 
and the value for the strong coupling do not change much in the present variant of the fit.
We observe consistency within the $\pm 1 \sigma$ statistical error. 
For illustration we display the $\pm 1 \sigma$ band of absolute uncertainties for the (non-strange) light-quark 
(Fig.~\ref{fig:lq-pdfs} left) and the gluon PDFs (Fig.~\ref{fig:lq-pdfs} right) 
at the starting scale $\mu = 3$~GeV compared to ABKM~\cite{Alekhin:2009ni} and we observe good agreement.
The PDFs in Fig.~\ref{fig:lq-pdfs} result from a fit, 
where we have additionally constrained the charm-quark mass by the PDG value of Eq.~(\ref{eq:mcpdg}).
This results in $m_c(m_c) = 1.18 \pm 0.06 ({\rm exp}) \pm 0.03 ({\rm th})$ at NNLO$_{\rm approx}$ 
in very good consistency with Eqs.~(\ref{eq:mcpdg}) and (\ref{eq:mbarcnnlo}).
For bottom the value $m_b(m_b) = 4.19 \pm 0.12$ with a symmetric error has been used.
Since our analysis involves fixed-target data from CCFR/NuTeV, 
we have also paid particular attention to the strange-quark PDF and a potentially asymmetric strange sea.
However, we find no indication. The total integrated asymmetry 
with the $\pm 1 \sigma$ statistical uncertainty at the scale 
$\mu^2 = 20$~GeV$^2$ is obtained as
\begin{eqnarray}
  \label{eq:s-sbar}
  \int_0^1\, dx\, x\, \left(s(x,\mu) - {\bar s}(x,\mu)\right) = 0.0011(9)
  \, ,  
\end{eqnarray}
which is consistent with previous results~\cite{Alekhin:2008mb}.

In Fig.~\ref{fig:zeus} we confront data from NC heavy-quark DIS 
with the resulting predictions using running masses for charm and bottom, $m_c(m_c) = 1.18$ and $m_b(m_b) = 4.19$.
The ZEUS data~\cite{Chekanov:2003rb} displayed in Fig.~\ref{fig:zeus} has not been used in the fit. 
At the smallest values of $x$ and $Q$ in Fig.~\ref{fig:zeus} the predictions 
rise monotonically with increasing orders of perturbative QCD, thus improving agreement with the data. 
In this region the value of $F_2^p$ is sensitive 
to the coefficient functions for small $\beta$, where the threshold approximation 
NNLO$_{\rm approx}$ is valid and can be considered as a good approximation to the full (yet unknown) NNLO result for $F_2^p$.
As can be seen in Fig.~\ref{fig:zeus}, with the chosen value for charm ($m_c(m_c) = 1.18$) 
our predictions are still slightly below data for at small values of $x$ and $Q$.
At large values of $x$ and $Q$ the slope of $F_2^p$ flattens in $Q$, particularly for higher values of $x$, 
and the agreement with data is still very good.
Future comparisons to high precision NC heavy-quark DIS data from the Run II of HERA will be interesting.

\bigskip

A couple of interesting remarks can be made at this point.
First of all, the experimental input to the PDG determinations 
of $m_c$ and $m_b$ originates entirely from $e^+e^-$-collisions or $B$-decays  
(see e.g.~\cite{Bodenstein:2010qx,Chetyrkin:2010ic} for recent analyses of $e^+e^-$-annihilation data with QCD sum rules).
While the use of short distance masses is by now fairly standard in cross section predictions 
for those processes, it has not been used much in phenomenology at hadron colliders, 
although it is well known that the pole mass is plagued by large intrinsic ambiguities.
However, with the increasing experimental precision of hadron collider data, 
there is a clear need to provide perturbatively stable theory predictions and
to use well-motivated definitions of fundamental quantities like the mass parameter.

The \MSbar masses of Eqs.~(\ref{eq:mbarcnlo}) and (\ref{eq:mbarcnnlo}) 
provide the first theoretically consistent determinations of these fundamental
parameters in heavy-quark DIS, a process governed by the exchange of space-like bosons.
So the agreement with the PDG value \cite{Nakamura:2010pdg} within the quoted accuracy 
is very reassuring also with respect to the different underlying (space-like) kinematics.
Previously, charm mass determinations in heavy-quark DIS have been performed
by CHARM II~\cite{Vilain:1998uw} and NOMAD~\cite{Astier:2000us}. 
They have extracted a value of $m_c$ from di-muon events in neutrino-nucleon DIS 
with a rather large uncertainty and based on a LO QCD analysis only. 
Also from the CCFR/NuTeV data for neutrino-nucleon DIS a value for the charm
mass has been extracted~\cite{Seligman:1997fe}, which however is effectively LO only, 
as far as the mass dependence of the cross section is concerned.
None of these determinations enters the world average as quoted by the PDG.

Another interesting issue concerns the heavy-quark PDFs. 
These are needed at high-energy hadron colliders for hard scattering processes at scales $Q \gg m_c, m_b$, 
where a FFNS with effectively $n_f = 4$ or $n_f = 5$ light flavors is the appropriate description.
The PDFs for charm- and bottom-quarks in 4- and 5-flavor schemes can be generated from the ones obtained 
in a 3-flavor FFNS as convolutions of the gluon and flavor-singlet distributions with massive operator matrix elements. 
Through the explicit mass dependence of the latter the uncertainty on heavy-quark PDFs 
is directly related to the accuracy of the quark mass parameter.
Preliminary studies have shown that precision input for the values of charm and bottom masses 
can greatly improve the accuracy of charm- and bottom-quark PDFs.
Future studies will be devoted to an in-depth investigation of the transitions from 3- to 4- to 5-flavor FFNS 
with the running mass scheme~\cite{abm:2010}.

Finally, let us address the implications for LHC processes, which are quite clear. 
In global PDF fits, the predicted rate for $W^\pm$- and $Z$-boson production 
is very sensitive to the chosen pole mass value for charm in schemes with four or five active light flavors.
At the $\sqrt{S} =7$~TeV LHC, for instance, Ref.~\cite{Martin:2010db}, has reported shifts 
in the total cross sections for $W^\pm$- and $Z$-boson production 
of more than $2\%$ resulting from small variations of $\rm 0.15$~GeV 
in the pole mass value for $m_c$ around a central value (see Tab.~\ref{tab:polemassvalues}).
For the $\sqrt{S} =14$~TeV LHC, these uncertainties increase to more than $3\%$.
These findings are worrisome as they potentially invalidate precision
predictions for these important high precision measurements of Standard Model processes.

Fortunately, as we have demonstrated, these uncertainties can be almost
entirely eliminated by adopting the running mass.
By using e.g. the very precise world averages as constraints in global fits, the errors on $m_c$ and $m_b$ can be greatly reduced. 
Moreover, these errors can be directly propagated to the PDF uncertainties with no need for additional assumptions.
In this manner, very accurate and precise predictions for $W^\pm$- and $Z$-boson production at the LHC are possible.

\section{Summary}
\label{sec:sum}

We have studied the production of heavy quarks in NC and CC DIS including higher order
radiative corrections in QCD adopting the running \MSbar mass and 
we have demonstrated the clear advantage of using this scheme.
The resulting theory predictions display an improved apparent convergence through NNLO$_{\rm approx}$ 
as compared to the result in the pole mass scheme, especially in the NC case.
Also the stability of the perturbative series with respect to scale variations is much greater.

We have used our results to determine for the first time the \MSbar mass for the charm-quark 
$m_c(m_c)$ 
based on first principles in QCD from a fit to DIS data for heavy-quark production.
The obtained value is consistent with the world average as published by the PDG 
and it provides complementary information on this fundamental parameter 
from hadronic processes with space-like kinematics.

We have shown that the use of \MSbar masses in heavy-quark DIS can also 
improve predictions for hadron colliders by eliminating sizable uncertainties in PDFs.
This underpins the need for global fits of PDFs to adopt the running mass scheme.
Current global fits of PDFs employ the pole mass scheme and small variations in the chosen value 
for the charm mass can easily amount to differences of a few per cent in the predicted 
cross sections of $W^\pm$- and $Z$-bosons at LHC energies.
With a physically well motivated short distance mass these uncertainties can be eliminated to a large extent. 
Progress in this direction will be reported elsewhere~\cite{abm:2010}.

\bigskip

The numerical code for the computation of structure functions for heavy-quark production in deep-inelastic scattering 
with a running mass is publicly available for download from~\cite{openqcdrad:2010} or from the authors upon request.

\subsection*{Acknowledgments}
We are thankful to J.~Bl\"umlein for discussions. 
This work has been supported by Helmholtz Gemeinschaft under contract VH-HA-101 ({\it Alliance Physics at the Terascale}). 
S.A. also acknowledges partial support from the Russian Foundation for Basic Research under contract RFFI 08-02-91024 CERN\_a.

\newpage
\newpage

\newpage

\end{document}